\def\etal{{\it et al. }}

 \documentclass[final,3p,times]{elsarticle}
\usepackage{amssymb}
\usepackage{amsmath}
\usepackage{float}
\usepackage{subfigure}
\usepackage{color}
\usepackage{adjustbox}
\usepackage{color, colortbl}
\definecolor{Gray}{gray}{0.75}
\definecolor{LightCyan}{rgb}{0.50,1,1}

\usepackage[section]{placeins}
\usepackage{amsfonts}
\usepackage{graphicx}
\usepackage{lipsum}
\usepackage{multirow}
\usepackage{booktabs}
\usepackage[table,xcdraw]{xcolor}
\definecolor{cream}{RGB}{222,217,201}

\journal{Acta Materialia}

\begin {document}
\begin{frontmatter}

\title{Machine-learning enabled thermodynamic model for the design of new rare-earth compounds}

\author{P. Singh$^{a,*}$, T. Del Rose$^{a,b}$, G. Vazquez$^c$, R. Arroyave$^{c,d}$, Y. Mudryk$^a$}
\address{$^a$Ames Laboratory, U.S. Department of Energy, Iowa State University, Ames, Iowa 50011 USA}
\address{$^b$Materials Science \& Engineering, Iowa State University, Ames, Iowa 50011 USA}
\address{$^c$Department of Materials Science and Engineering, Texas A\&M University, College Station, TX 77843, USA}
\address{$^d$Department of Mechanical Engineering, Texas A\&M University, College Station, TX 77843, USA}

\cortext[mycorrespondingauthor]{Corresponding authors email:\textrm{psingh84@ameslab.gov}} 

%%%%%%%%%%%%%%%%%%%%%%%%%%%%%%%%%%%%%%%    
\begin{abstract}
We employ a descriptor based machine-learning approach to assess the effect of chemical alloying on formation-enthalpy of rare-earth intermetallics. Application of machine-learning approaches in rare-earth intermetallic design have been sparse due to  limited availability of reliable datasets. In this work, we developed an `in-house' rare-earth database with more than 600$+$ compounds, each entry was populated with formation enthalpy and related atomic features using high-throughput density-functional theory (DFT). The SISSO (sure independence screening and sparsifying operator) based machine-learning method with meaningful atomic features was used for training and testing the formation enthalpies of rare earth compounds. The complex lattice {function coupled with the machine-learning model was used to explore the effect of transition metal alloying on the energy stability of Ce based cubic Laves phases (MgCu$_{2}$ type).} The SISSO predictions show good agreement with high-fidelity DFT calculations and {X-ray powder diffraction measurements.} Our study provides quantitative guidance for compositional considerations within a machine-learning model and discovering new metastable materials. {The electronic-structure of Ce-Fe-Cu based compound was also analyzed in-depth to understand the electronic origin of phase stability.} The interpretable analytical models in combination with density-functional theory and experiments provide a fast and reliable design guide for discovering technologically useful materials.
\end{abstract}

%%%%%%%%%%%%%%%%%%%%%%%%%%%%%%%%%%%%%%


\end{frontmatter}

%%%%%%%%%%%%%%%%%%%%%%%%%%%%%%%%%%%%%
\section*{Introduction and Background}
%%%%%%%%%%%%%%%%%%%%%%%%%%%%%%%%%%%%%

Rare earths find uses in many applications due to their vast span of distinctive physical and chemical properties.  Well-known applications include wind turbines, hybrid and electric vehicles, solid state lighting, mobile devices, lasers and optical fibers, fuel cracking and other catalysts, all of which owe most of their core functionalities to 4$f$-electron compounds \cite{1,2,3,4,5,6,7,8,9,10,11,13,14}. A number of future technologies such as magnetic cooling \cite{15_1,18} rely on rare earths for both magnetocaloric materials and magnetic field sources. {Recently, a global ``push" for a greener, hydrogen economy brings attention towards solid-state hydrogen liquefaction technology, where rare earth alloys (in particular RT$_{2}$ compounds where T = Al, Ni, Co) have potential to play a vital role \cite{Park2017,Gschneidner1994,Numazawa2014,Castro2020}. However, the} discovery of rare earth compounds is often built upon serendipitous findings, as can be exemplified by hard rare-earth magnets \cite{19,20,21}. While such encounters will remain a part of future scientific progress in, there is a growing consensus that predictive science enabled by recent developments in theory and machine learning will lead the way for sustainable innovation \cite{Amato2019}. 

The lanthanides (mostly trivalent in nature with the exception of Eu and Yb, where a divalent state could be more stable, and Ce, which is often tetravalent) and their compounds have always been a fertile ground for informatics and prediction due to their peculiar chemistry. The concept of the "fraternal fifteen", imagined by K.A. Gschneidner, Jr., presented lanthanides as consecutive houses on the same street but with different number of "kids" or 4$f$ electrons \cite{22}.  When moving through the lanthanides, the ``houses" (sub-orbitals, e.g, 4$f$) are filled with varying numbers of kids ($f-$electrons), while the house's exterior (outer shells) remain relatively unchanged ($\left[Xe\right]$4$f^{n}${\bf 5$d^{1}$6$s^{2}$} or, in some cases, $\left[Xe\right]$4$f^{n}${\bf 6$s^{2}$}). A famous example constitutes a generalized phase diagram for the trivalent R$_{z}$R$'_{1-z}$ intra-lanthanide alloys, which predicts crystal structure and melting temperature of compositions with any specific z value \cite{23}. A modified version of this diagram was constructed to predict phase boundaries of intra lanthanide alloys at pressures up to 25 GPa \cite{24}. This chemical similarity together with a well-established evolution of atomic radii, known as lanthanide contraction, and established rules of some physical behaviors (e.g., de Gennes rule for magnetism) provide a fertile ground for systematic generalization of respective physical and chemical behaviors, suitable for predictive science. 

Machine learning models have proven to be useful in identifying crystallographic information of ternary equiatomic rare earth compounds \cite{25, 26}. These ML efforts, however, are often limited in their ability to describe true thermodynamic behavior of  new rare earth compounds or compositions. Nonetheless, despite the importance of rare-earths based materials and their solid background, modern machine learning and artificial intelligence tools for the targeted investigation of rare earths remain sparse due to a lack of reliable databases with sufficient number of entries required for model training. Naturally, this emphasizes the need of models that can interpolate data from sparse databases. The SISSO ($S$ure $I$ndependence $S$creening and $S$parsifying $O$perator) is one such machine-learning method \cite{Ouyang2018,Ouyang2019} that returns very  accurate predictions with limited information~\cite{Bartel2018, Bartel2019, Sauceda2021,Singh2021_2}. The readiness of SISSO-based analytical models using regression and classification makes the model interpretation easier compared to other machine learning methods that face the non-trivial question on the representation of material design space. The SISSO models return the best analytical descriptors trained with limited information for a target property, e.g., thermodynamic stability or formation enthalpy, from a vast feature space constructed from an operator set (e.g., $`+'$, $`-'$, $`exp'$, $`log' $ etc.) and an initial primary feature set (e.g., atomic-size, valence-electron count, Allen electronegativity etc.)\cite{Ouyang2018}.

In this work, we present a rapid assessment of alloying effects on the thermodynamic stability of REX$_{2}$ (RE=Rare-earth; X=transition-metals) type rare-earth intermetallics using machine-learning (SISSO) trained inexpensive analytical models. The machine-learning model was trained over the density-functional theory (DFT) generated {\it in-house} database with 600$+$ entries. During the training of the model, a total of 26 material descriptors were shortlisted from an extensive pool of 89 features (see supplement) based on their fundamental relevance and statistical correlation with formation enthalpy (see Fig.~S1). The Ce-Fe-Cu rare-earth compounds, in this work, were chosen for the model validation. The potential application of Ce-based compounds as high-performance permanent (hard) magnets make them an ideal candidate for phase stability and electronic-structure analysis. The SISSO trained three-dimensional (3D) descriptor was used to analyze the alloying effect on the phase stability of RE(TM1$_{z}$TM2$_{1-z}$)$_{2}$ compounds with cubic structure (MgCu$_{2}$ type), where RE=Ce; TM1=Mn/Fe/Co/Ni; and TM2=Pd/Pt/Re/Rh, and z is the variation in transition metal composition. The phase stability trends are validated through very careful synthesis of Ce-Fe-Cu based rare-earth compounds, i.e., Ce(Fe$_{z}$Cu$_{1-z}$)$_{2}$ at z(Fe)= 25, 50 and 75 at.\%Fe, followed by detailed X-Ray powder diffraction analysis. {A comprehensive analysis of chemical alloying on the electronic-structure (the band-structure and Fermi-surface) of Ce-Fe-Cu compounds in cubic (MgCu$_{2}$) and orthorhombic (KHg$_{2}$) phase is also presented to understand the quantum mechanical origin of phase stability.} The change in band-structure with Cu strongly correlates with the predicted phase stability. Our work provides a logical next step in the informed discovery of new rare earth compounds by integrating machine learning models with high throughput DFT calculations. This further emphasizes the need of computationally inexpensive and interpretable models for the accelerated discovery of complex alloys.

 %%%%%%%%%%%%%%%%%%%%%%%%%%%%%%%%%%%%%
\section*{Methods}
%%%%%%%%%%%%%%%%%%%%%%%%%%%%%%%%%%%%%
\subsection*{Database generation}

{\it Structure considerations for database generation:}~Ames laboratory internal Rare-Earth Information Center (RIC) database, containing over 100,000 references to rare earth materials published before the early 2000s, was scraped for compositional information. Of the resulting 59,000$+$ potential compositions, select rare earth material classes (REX$_{2}$ , REX, RE$_{2}$X, etc) were used to query the Springer Materials database for Crystallographic Information Files (CIF).  {The CIFs containing high-temperature and high-pressure structures were used for testing the physical accuracy of our models, while the ground state CIFs (verified through DFT calculations of formation enthalpy) were used in both training and testing.In case of multiple crystal structures, the low temperature phases at ambient pressure were chosen as the ground state to populate the database.} 

{\it High-throughput density-functional theory calculation:}~First-principles density functional theory (DFT) as implemented in Vienna Ab-initio Simulation Package (VASP) \cite{VASP1,VASP2}  was used for high-throughput generation of structural (e.g., lattice constants, volume etc.) and electronic properties (such as formation enthalpy and valence-electron count). The generalized gradient approximation of Perdew, Burke and Ernzerhof (PBE) was employed in all calculations \cite{PerdewPBE}. {The DFT+U \cite{Dudarev1998} is an alternate scheme for materials with localized $d-$ and $f-$electrons. However, the non transferability of parameter U across compounds and its arbitrary nature, where the choice of U can strongly influence  the observables \cite{Loschen2007}, makes the high-throughput use of DFT+U approach impossible. Franchini et al.~\cite{Franchini2007} have shown that DFT+U method introduces an average (minimum) relative error of about 4\% in calculation of heat of formation in binary-manganese oxides, where a single U value was used. Additionally, S\"oderling et al.~\cite{Soderlind2014} also established the effectiveness of GGA functionals in describing the properties of rare-earth metals. The hybrid functionals are less relevant to this discussion for a couple of reasons: (i) most hybrid-functionals are one-shot calculations (no self-consistency), and (ii) they are computationally very demanding, therefore, not useful for machine-learning approaches. Furthermore, the rationale to choose GGA (PBE) over LDA, meta-GGA \cite{Singh2013,Singh2016} or hybrid functionals \cite{Becke1993} was inspired by the work of Giese and York \cite{Giese2010} which highlights the advantage of GGA functionals. Therefore, PBE (GGA) was used in all our calculations \cite{PerdewPBE}.}

Based on energy and $k$-mesh convergence tests, we set up high-energy plane-wave cutoff  of 520 eV both in relaxation and charge self-consistency. Each crystal structure was fully relaxed (volume $+$ atomic-positions) with high convergence criterion for force ($10^{-3}$ eV\AA) and energy ($10^{-6}$ eV). {Full relaxation and charge self-consistency were done using 2$\times$2$\times$2 to 8$\times$8$\times$8 depending on total number of atoms per cell (the inverse relation between k-space and r-space helps to choose smaller $k-$mesh for larger cells and vis-a-vis). The $\Gamma$-centered Monkhorst-Pack $k$-mesh was used for Brillouin zone integration during structural-optimization and charge self-consistency \cite{MP1976}. The shift away from $\Gamma$ was originally meant to reduce the size of the $k-$mesh. In metallic compounds, $\Gamma-$centered sampling was chosen as it is often a better choice for its faster-convergence than the shifted grid \cite{Baldereschi1972}.}

{\it Formation enthalpy estimate:}~The formation enthalpy (E$_{form}$) of rare-earth based REX$_{2}$ systems was estimated using the formula E$_{form}$=E$_{total}^{REX_{2}} - \sum_{i}$N$_{i}$ E$_{i}$  where E$_{total}^{REX_{2}}$ is the total energy per unit cell of the intermetallic rare-earth compounds per unit cell, N$_{i}$ and E$_{i}$ are the  number of atoms and ground state total energy of element `i' (e.g., Fe in body-centered cubic, and/or Ce in face-centered cubic).

\subsection*{SISSO- model-training and accuracy}

The SISSO-based ML approach combines regression with compressed sensing to identify high-order analytical models, for example, 1D, 2D, 3D, 4D or even higher dimensional interpretable descriptors \cite{Ouyang2018,Ouyang2019}. The SISSO models work in a hierarchical order where this first creates a new feature space of varying complexity through algebraic operations ($+, - , \times,e^{x},e^{-x},()^{-1},()^{2},()^3,()^6,\sqrt{ },log$) on primary feature sets (variables). This allows us to create sufficiently large feature space. All features generated through the combination of the primary features and operators were added recursively to the feature set. The sparse regression method then filters out significant features from an originally large feature space based on their frequency in descriptor generation. We have performed the SISSO training and testing using 5- and 10-fold cross-validation. 

{The SISSO models were trained on all lanthanides (from La to Lu except Pm) and transition-metals (VIIA-XIIA), as well as $I$A , IIA, IIIA (B-Ga), IVA (C-Sn), VA (N-Sb), VIA (O-Te), VIIA (F-I) group elements. The use of such diverse dataset makes the model generalized enough to make predictions of any rare-earth based REX$_{2}$ compounds. Since the group IVB (Ti, Zr, Hf), VB (V, Nb, Ta), and VIB (Cr, Mo, W) transition metals were not included in the model training, so the confidence in prediction may be low for  REX$_{2}$ compounds with X=IVB, VB, VIB group elements.}

%%%
\subsection*{Weighted sublattice description}
In general, the featurization of ML-based models involves composition based vectors for application to disorder materials. Evidently, the number of features grows with increasing alloy complexity, therefore, it would be more convenient if we separate the alloy features by the role of constituent elements. We used the idea of sublattice description to distinguish the lattice sites by rare earth (RE) and transition metals (X) in REX$_{2}$. An average value was assigned to each Wyckoff site, i.e., 8$a$ (RE) and 16$d$ (X), where the transition-metals are twice the amount of rare-earths. A weight feature was created (e.g., `plus' and `plus-2';  `minus' and `minus-2') to calculate the contribution of both the sites according to their weights. An average over the lattice-cell was calculated as an weighted average  and weighted standard deviation over the whole cell (see supplemental Eq.~S1-S3).
%%%

\subsection*{Experimental details} 

Three Ce(Fe$_{z}$Cu$_{1-z}$)$_{2}$ samples with z (Fe)=25, 50, 75 at.\% were prepared to verify the mixability of Fe and Cu in the  Ce(Fe$_{z}$Cu$_{1-z}$)$_{2}$  solid solution. The samples were arc-melted using the high-purity metals: Fe and Cu were at least 99.95 wt.\% pure; Ce (at least 99.9 wt.\% pure) was provided by Materials Preparation Center of Ames Laboratory. The samples were wrapped in Ta foil, sealed in evacuated quartz ampoules back-filled with He gas, and annealed at 800$^{o}$C for 2 weeks. The samples were slowly cooled in the furnace and analyzed using X-ray powder diffraction (XRPD) and electron microscopy (SEM/EDX). The SEM analysis revealed strong surface oxidation, however, the EDX confirmed that the chemical composition of the prepared samples matches the nominal one. The phase analyses of the XRPD data was performed using Rietica software \cite{Hunterxxx}.

\section*{Results and Discussion}

\subsection*{Formation enthalpy - training, model-generation, cross-validation, and feature analysis}

\noindent
{\it Training, testing and descriptor analysis:~}  Machine learning and data analytics can accelerate materials design and discovery through the use of descriptors \cite{17_2,18_2,19_2,20_2,21_2,22_2,23_2,24_2}. Ouyang \etal~\cite{Ouyang2018,Ouyang2019} has shown the usefulness of symbolic regression in SISSO that allow us to develop analytical descriptors for target properties \cite{25_2,26_2,27_2}.  Despite descriptors becoming a standard approach for material discovery, literature remains sparse on databases containing thermodynamic stability information  (e.g., formation enthalpy; E$_{form}$) of rare-earth based inorganic crystalline solids. 

The E$_{form}$ is a fundamental material property and an important indicator of thermodynamic stability. The `in-house' Ames Lab Rare-Earth Information Center (RIC) database (`Ames Lab RIC 2.0') was used to train the SISSO-based formation enthalpy descriptor. An analytical 3 dimensional (3D) formation enthalpy descriptor from SISSO model training was presented in Eq.~\ref{Eq_eform}: 

\begin{eqnarray}\label{Eq_eform}
E_{form} =  -158.157\times{10^{-3}}\times{log\left[r_{c}\times{\rho_{avg}}\times{\chi_{diff}^{Pauling}}\right]}+ 18.585\times {10^{-3}}\times abs\left[\frac{1}{\rho_{avg}-\rho_{diff}} \right] \\ \nonumber + 3.28\times{10^{-3}}\times(group_{avg}-group_{diff})\times(r_{diff}^{Zunger}-r_{diff}^{Miracle});
\end{eqnarray}
\noindent
where, $r_{c}$ is the covalent-radii (\AA), $\rho_{avg}$, and $group_{avg}$ are averaged with stoichiometrically weighted mean and $\rho_{diff}$, $r_{diff}$, $\chi_{diff}$, and $group_{diff}$ are stoichiometrically weighted harmonic mean, see Eq.~S1-S3 in the supplement. The `in-house' rare-earth database was randomly divided (80$:$20 ratio) for training$:$testing of the SISSO based machine-learning model. The Eq.~\ref{Eq_eform} will allow the high-throughput determination of thermodynamic stability of the specific crystal phase and compositions that could be synthesized experimentally. The analytically constructed descriptor in Eq.~\ref{Eq_eform} also conserves the unit of feature sets within SISSO \cite{Ouyang2018}.

\begin{figure}[H]
    \centering
    \includegraphics[width=1\columnwidth]{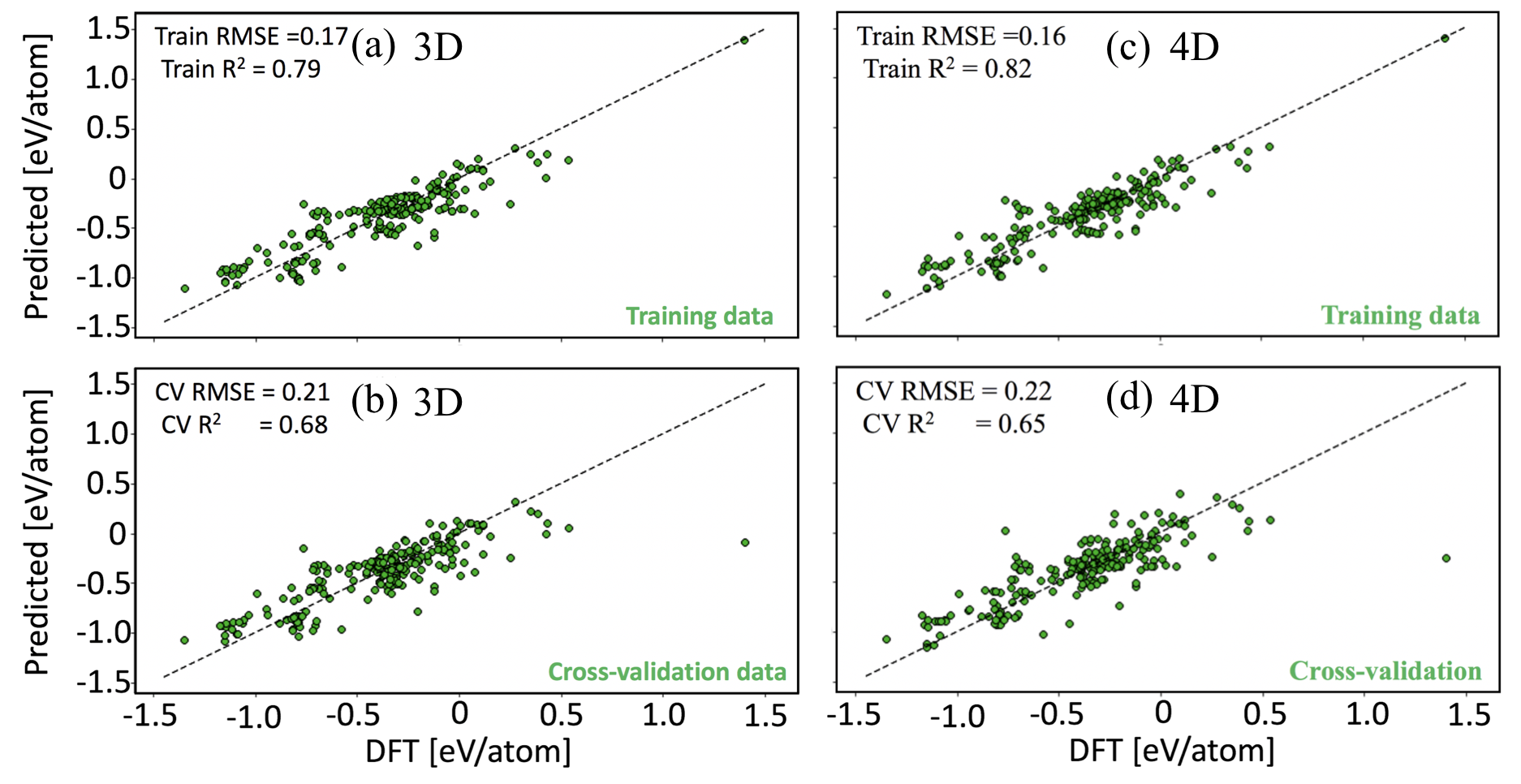}
    \caption{Predicted (descriptor in Eq.~\ref{Eq_eform}) vs actual (DFT) formation enthalpies from (a,b) 3D, and (c,d) 4D descriptor.}
    \label{fig2a}
\end{figure}

The sensitivity of 3D descriptors on training data was tested by performing 5 (10)-fold cross validation of Boufounos \etal~ \cite{Boufounos2007}, where  `Ames Lab RIC 2.0' was equally and randomly divided into 5 (10) sets. Following this, one out of 5 (10) set was left out as a test set and remaining were included in the training. {The cross validation was iterated 5 (10) times so that each part was used as a test once during the training. The cross-validation often helps to determine overfitting or underfitting of the models. Notably, dimensionality (either too large or too small) is the only source of overfitting in SISSO \cite{Ouyang2018}.} In 3D vs 4D descriptor comparison, we found no visible improvements in model accuracy as shown by root-mean squared error (RMSE; 0.17 eV-atom$^{-1}$ (3D) vs 0.16 eV-atom$^{-1}$ (4D)) and coefficient of determination (R$^{2}$; 0.79 (3D) vs 0.82 (4D)) in Fig.~\ref{fig2a}a (the error analysis see Eq. S4-S5). Therefore, in this work, the 3D descriptor was was used throughout for thermodynamic stability analysis of REX$_{2}$ compounds. The observation about dimensionality was found in good agreement with Ouyang \etal~\cite{Ouyang2018}, where the disadvantages of using low or higher order descriptors were clearly outlined. The computational cost is also a critical issue to keep in mind while choosing model dimensionality; therefore, it becomes critical to limit unnecessary features, which is only possible with the choice of optimal descriptor dimension.

Going back to the interpretation of formation-energy descriptor, a strong correlation among features such as electronegativity, density and atomic-radii with E$_{form}$  in Eq.~\ref{Eq_eform} makes sense. The atomic stability, also discussed in literature \cite{Rahm2021}, could be rationalized based on atomic radii and electronegativity, which has previously shown usefulness in describing bond character and compound formation \cite{Pauling1960,Rahm2016}.  For example, the electronic distribution around a nucleus in an atom determines the atomic-size, i.e., electrons closer to nuclei make a tightly bound system that eventually enhances the electronegativity \cite{Allen1989}. The increased screening by inner electrons moving down the group in the periodic table decreases the attraction force on the valence electrons, further balancing out the bonding with the nucleus. The change in electron screening reduces the electronegativity and increases the atomic-radii of the elements.

\begin{figure*}[t]
    \centering
    \includegraphics[width=1\textwidth]{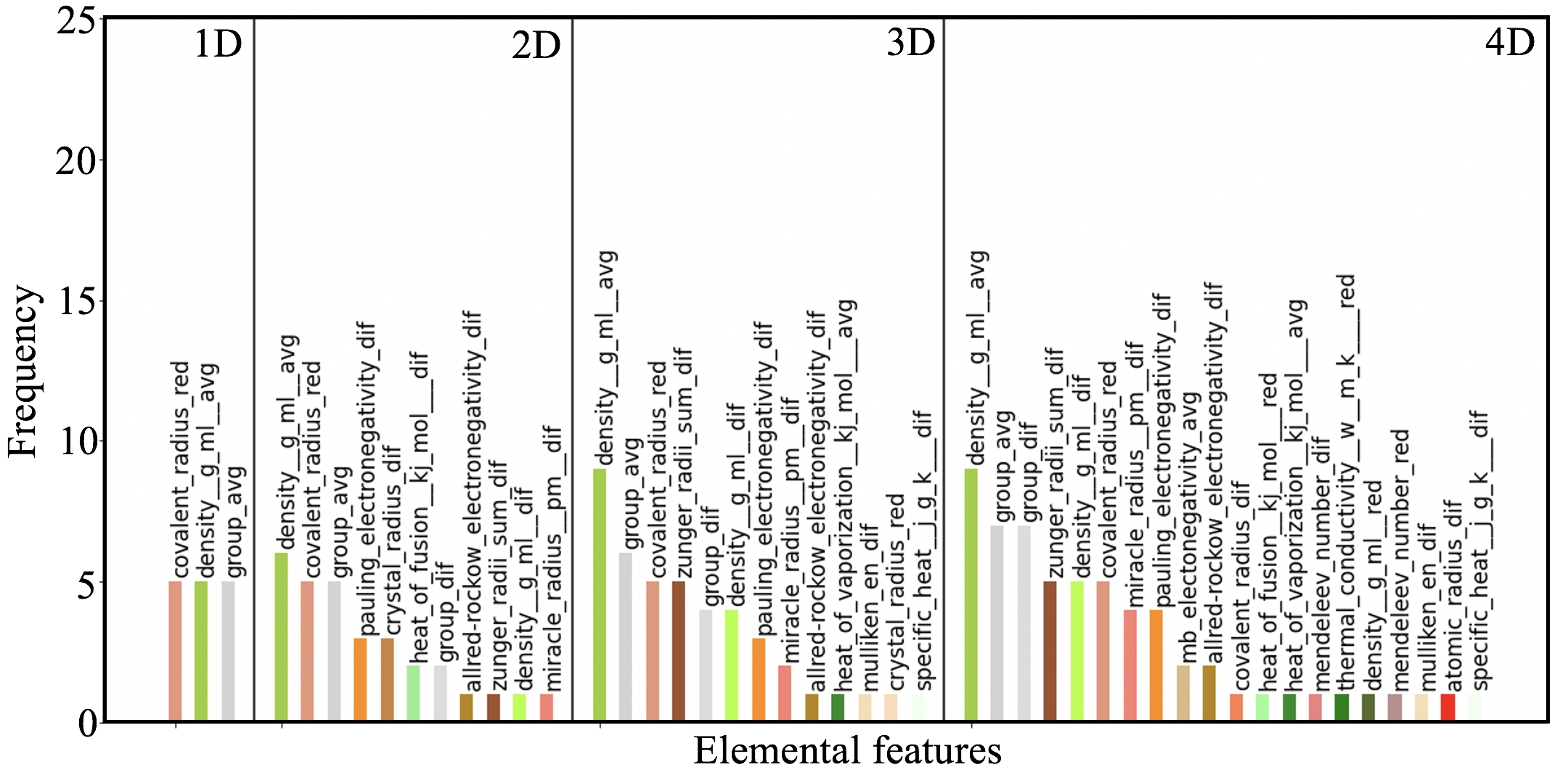}
    \caption{(Color online). Rank of features by appearance in 3D descriptor described by Eq.~\ref{Eq_eform}.}
    \label{fig:correlas}
\end{figure*}

%\noindent
{\it Feature analysis:~} The feature complexity was found to increase with increasing model dimensionality as shown by the large feature sets in higher order feature spaces in Fig.~\ref{fig:correlas}. The Sure Independence Screening (SIS) creates subspaces of strongly correlated features for the property of interest, in our case formation-enthalpy, from an exponentially large feature set of cardinality $\approx{10^{10}}$. The correlated features in the subspace were operated on by an $l_{0}-$norm regularized minimization Sparsifying Operator (SO ($l_{0}$)) to help find the best descriptor. The $l_{0}$-norm of a vector is the number of its non-zero components. The number of terms in the descriptor space depends on dimensionality of the subset space and can be mathematically represented by n-combination (where n represents the dimensionality, i.e., $1D$, $2D$, and $3D$), for example, the 3D descriptor chooses 3 set of features from an exponentially large feature space, i.e., $\left ( {3000 \atop 3} \right ) \sim 4.5 \times 10^9$. 

The frequency of features in the 3D descriptor space for a five-fold cross-validation maximizes at 10---see Fig.~\ref{fig:correlas}. The first eight most frequent features were selected by the descriptor without any exception to describe the $E_{form}$ in Eq.~\ref{Eq_eform}. The most frequent features are those generated through the stoichiometrically-weighted mean (e.g., $\rho_{avg}$), stoichiometrically-weighted harmonic mean (e.g., $r_{zunger}^{diff}$), and stoichiometrically-weighted mean difference  (e.g., $r_{covalent}^{red}$) \cite{Bartel2018}. Here, we stick to a 3D descriptor space as  changes in features may cause over-fitting in top performing higher dimensional descriptors \cite{Ouyang2018}. 
The feature frequency in Fig.~\ref{fig:correlas} was found near constant for 3D descriptors (4-to-10 features) and performs well both in training and cross-validation. The features in the final descriptor trained on the full dataset also follow the same trend in choosing the most recurring features in cross-validation. Notably, the 3D descriptor in Eq.~\ref{Eq_eform} shows higher confidence in top features in Fig.~\ref{fig:correlas}  such as covalent radii ($r_c$), average density ($\rho_{avg}$), Pauling electronegativity ($\chi_{Pauling}$), group average ($group_{avg}$), and group difference ($group_{diff}$). We also found that  the higher frequency of selected features appearing in each run improves the accuracy of the model and makes the descriptor acceptable for phase stability prediction. 

As pointed out by Schmidt \etal~\cite{Schmidt2019}, ideally, descriptors should be uncorrelated, as an abundant number of correlated features can hinder both accuracy and efficiency of the model. This suggests that one needs further feature selection to circumvent the negative effect of dimensionality \cite{Bellman2015}, simplify the models, and improve their interpretability as well as training efficiency~\cite{Schmidt2019}. As discussed by Ouyang \etal~\cite{Ouyang2018}, the analytical nature of the SISSO model helps to circumvent the effect of dimensionality.

\section*{Thermodynamic stability assessment of MgCu$_{2}$ type Ce compounds using the E$_{form}$ descriptor}

\subsection*{Background and crystal symmetry of cubic rare-earth Laves phases (MgCu$_{2}$ type)}
 
Of the 1,200$+$ Laves phases known today, about 60\% contain rare earths, and nearly $\sim$200 of those are binaries. Their excellent chemical and physical properties give them a unique advantage for application as functional materials \cite{Gschneidner2006}. The RE(TM)$_{2}$ Laves phase compounds are the closely related intermetallics that crystallizes in three different strucrtures such as MgCu$_{2}$ (C15), MgZn$_{2}$ (C14), and MgNi$_{2} $ (C36). 
Recently, some of the non rare-earth $C14$ Laves phases such as (Hf$_{1-z}$Ta$_{z})$Fe$_{2}$, (Sc$_{1-z}$Ti$_{z}$)Fe$_{2}$ and (Zr$_{1-z}$Nb$_{z}$)Fe$_{2}$ were found with interesting compositional and temperature dependence of their magnetic properties \cite{Nishihara1983,Nishihara1984,Yamada1984}. Similar effects were found in rare-earths such as Ce$_{1-z}$(Ru$_{z}$Fe$_{1-z}$)$_{2}$ with C15 crystal phase \cite{Hilscher,Roy2004}. Despite the intriguing nature of rare-earth based compounds, the alloying effects on the thermodynamic stability was not systematically assessed either with machine-learning or high-throughput DFT methods.

\begin{figure}[htp]
\centering
\includegraphics[width=0.6\textwidth]{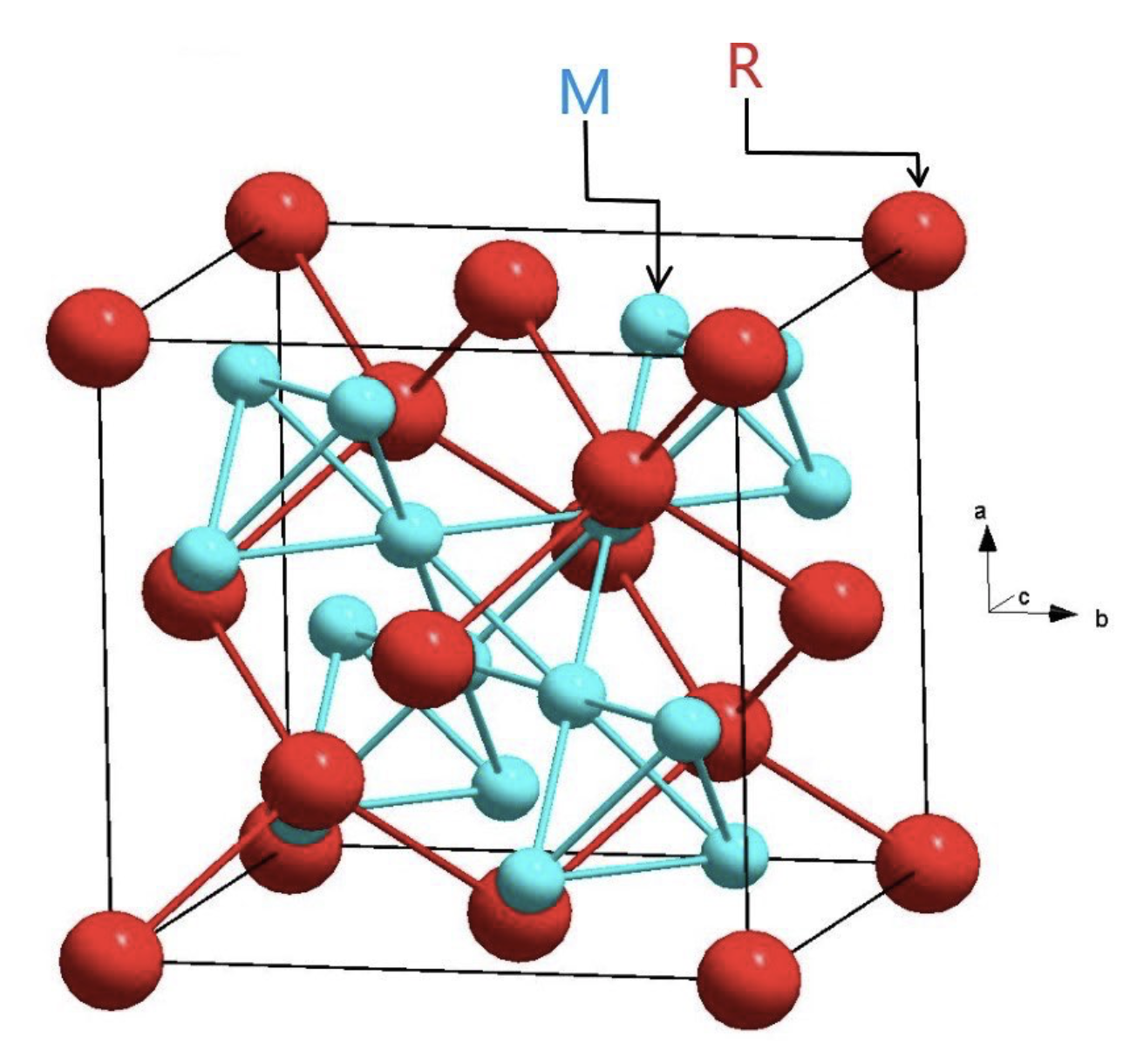}
\caption{(Color Online) Crystal structure of the MgCu$_{2}$-type structure.  The large (red) atoms represent the Rare Earth (Mg-site), and the smaller (cyan) atoms represent the transition metal (Cu-site).}
\label{C15}
\end{figure}

Here, we explored MgCu$_{2}$-type Laves phases as model systems as they are the most representative among the rare earth compounds~ \cite{Gschneidner2006}. The MgCu$_{2}$ crystal phase has face-centered cubic symmetry with $Fd-$3$m$ ($\#$227) space-group. The cubic MgCu$_{2}$ phase is alternatively represented by its common Strukturbericht designation C$15$. The Laves phases contains two (Z=2) formula units (f.u.) in primitive and eight (Z=8) f.u. in conventional cells. The Cu-site forms a tetrahedron at Wyckoff position 16c $\left[3/8,3/8,3/8\right]$ and Mg-site arranged on diamond cubic lattice with Wyckoff position 8b $\left[0,0,0\right]$, totaling 24 atoms per conventional unit cell as shown in Fig.~\ref{C15}. Similar to the pure metals, the rare-earth elements are arranged in a ABCABC packing sequence in the cubic MgCu$_{2}$-type structure.

Thermodynamic stability is a primary fundamental quantity that must be evaluated to determine synthesizability.  The main idea behind the exploration of the thermodynamic phase space of rare-earths compounds is to find out the rule of mixing that governs phase selection. Cerium-based binary and ternary intermetallics have been exhaustively explored in the past for their electronic properties~\cite{CeB6}, and large magnetic response \cite{Takabatake1987,Schmidt2005,Chevalier2006,Lenkewitz1996}. Ce-based alloys are fundamentally interesting for their rich physics \cite{1_3,2_3,3_3,4_3,5_3} that can further be tuned using chemical alloying, which may also help improve the physical properties. The polymorphism in Ce is also well established~\cite{Matara2013}, where different stable phases $\alpha$, $\alpha`$, $\beta$, $\gamma$, $\delta$ have been reported at different thermodynamic conditions~\cite{Johansson1973,King1970}. Other noted properties, e.g., superconductivity~\cite{Seyfarth2021,Luo1968}, heavy-fermion behavior~\cite{Onuki1987,Tang2007}, and complex magnetic properties~\cite{Streltsov2012,Mathur1998,Murani2005} due to varying oxidation states of Ce (nonmagnetic Ce$^{4+}$ ([Xe]4$f^{0}$), magnetic Ce$^{3+}$ ([Xe]4$f^{1}$)) make Cerium based system an interesting playground for new discoveries~\cite{Sun2017}.

\subsection*{Effects of chemical mixing of 3$d$ and 4$d$/$5d$ transition metals on phase stability}

\begin{figure}[htp]
    \centering
    \includegraphics[width=1\textwidth]{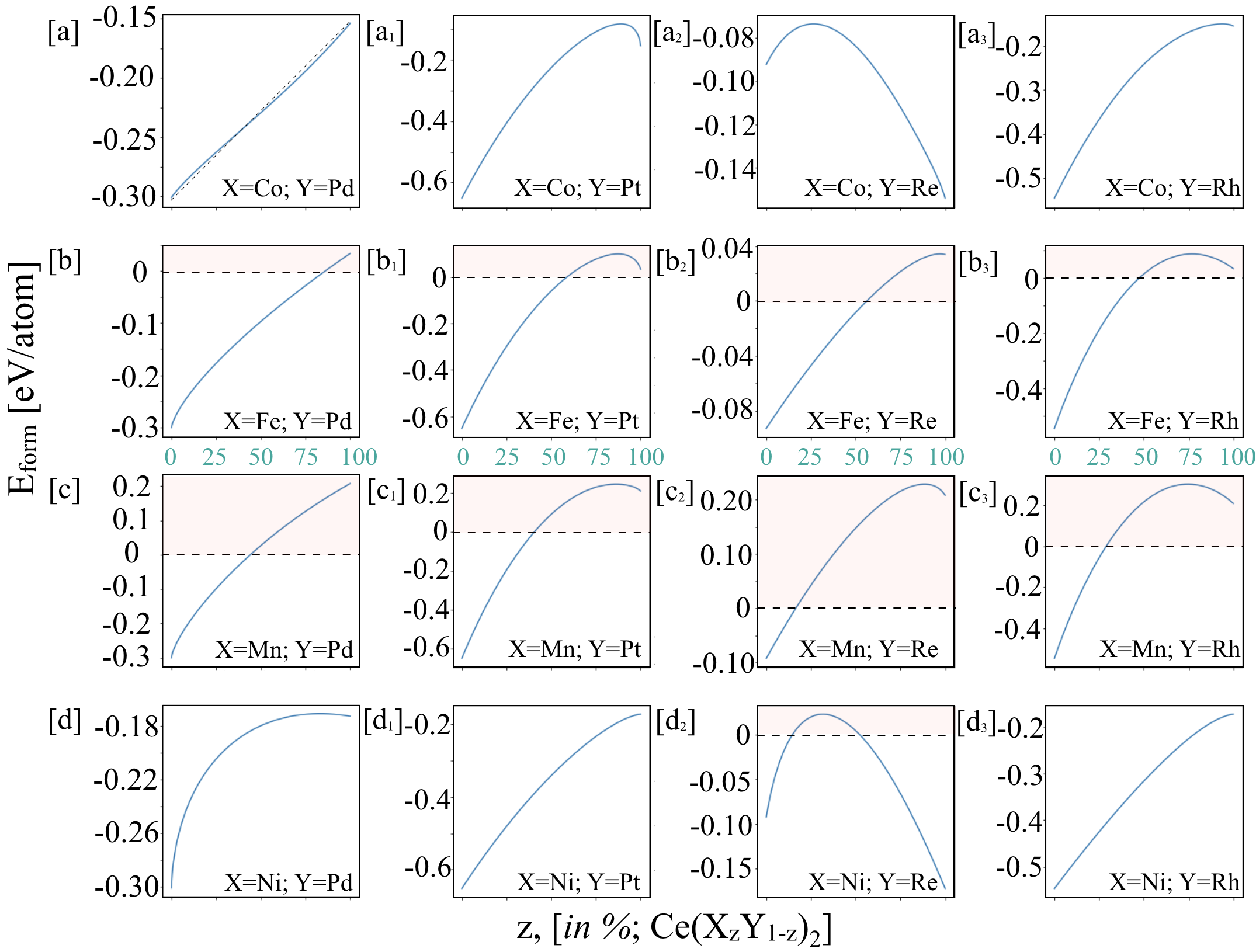}
    \caption{(Color Online) Thermodynamic stability of cerium based pseudo-binary compounds (a-a$_{3}$) Ce(Co$_{z}$TM2$_{1-z}$)$_{2}$, (b-b$_{3}$) Ce(Fe$_{z}$TM2$_{1-z}$)$_{2}$, (c-c$_{3}$) Ce(Mn$_{z}$TM2$_{1-z}$)$_{2}$, and  (d-d$_{3}$) Ce(Ni$_{z}$TM2$_{1-z}$)$_{2}$, where TM2=Pd/Pt/Re/Rh; and z is elemental composition in atomic-percent.}
    \label{Ce_stability}
\end{figure}

The thermodynamic stability analysis using experiments and direct {\it Ab-initio} (DFT) approaches is a time-consuming process \cite{Ivanov2013}. Here, we used a descriptor based analytical model as presented in Eq.~\ref{Eq_eform} for the high-throughput assessment of the phase stability of disordered rare-earths with MgCu$_{2}$ (cubic) crystal phase. We extensively explored the effect of transition metal alloying on the phase stability of pseudo-binary Ce(TM1$_{z}$TM2$_{1-z}$)$_{2}$ compounds, where the 3$d$ (TM1=Co/Fe/Mn/Ni) and  4$d$/5$d$ (TM2=Pd/Pt/Re/Rh) transition metals are alloyed at 16c Wycoff site in Fig.~\ref{C15} to understand their effect on thermodynamic stability. In Fig.~\ref{Ce_stability}a-a$_{3}$, we show the effect of Pd/Pt/Re/Rh alloying with Co on the phase stability of Ce-Co pseudo-binaries. Chemical mixing of Pd or Pt with Co in Fig.~\ref{Ce_stability}a\&a$_{1}$ shows improvement in the  phase stability the Ce-Co-Pd/Pt compound. However, the linear trend suggests that Co and Pd do not want to mix together rather they may prefer to phase separate but still coexist as two phase region of CeCo$_{2}$ and CePd$_{2}$. Similarly, the positive slope of phase stability in Fig.~\ref{Ce_stability}a$_{1}$ for Co-rich Co-Pt compound shows un-mixing of Pt at Co, i.e., Co energetically does not prefer to mix with Pt at transition metal site in Fig.~\ref{C15}.

In Fig.~\ref{Ce_stability}a$_{2}$, the strongly reduced (weak formation enthalpy) phase stability of Ce-Co compounds on alloying Co with Re agrees well with Co-Re binary phase diagram, where hexagonal phase becomes more stable at higher at.\%Re. Meanwhile, no experimental reports on cubic CeRe$_{2}$ further affirms the validity of our predictions \cite{Gorr2009}. The phase stability analysis indicates that CeRe$_{2}$ is very weakly stable, i.e., even if the compound forms it is either metastable or dynamically unstable in ambient conditions. Therefore, it needs to be looked with respect to other competing phases such as hexagonal MgNi$_{2}$ crystal type. {Notably, the ML predicted formation enthalpies for CeRe$_{2}$/Ce(CoRe)$_{2}$/CeCo$_{2}$ (-0.09/-0.085/-0.17 eV-atom$^{-1}$) show reasonable agreement with direct DFT calculations (-0.06/-0.11/-0.26 eV-atom$^{-1}$ ).} The jump in formation enthalpy at intermediate compositions (near 50at.\%Re) of Ce-Co-Re compound in Fig.~\ref{Ce_stability}a$_{2}$ is suggestive of structural transformation. Similar jump was also observed in volume near 50at.\%Re, which indicates towards possible phase change. Moreover, the existence of mixed valency in Ce, e.g., trivalent (+3) and tetravalent (+4), can also lead to the phase change in Ce-Fe compounds, for example, smaller Ce$^{4+}$ ion radii compared to Ce$^{3+}$ makes it easier to form the CeFe$_{2}$ phase \cite{Zhao2018}. Mixed valence state, and the fact that the ground state structure of CeFe$_{2}$ is not cubic, increases the challenge of accounting it into ML models, therefore, for simplicity, not considered. In Fig.~\ref{Ce_stability}a$_{3}$, we show the effect of alloying Rh-4$d$ (Y) on the phase stability  of Ce(Co$_{z}$TM2$_{1-z}$)$_{3}$, where Rh-4$d$ increases the alloy phase stability. Experiments also show that CeRh$_{2}$ stabilizes in cubic phase \cite{Palenzona1993}, which further confirms the robustness of our predictions.

In Fig.~\ref{Ce_stability}b-b$_{3}$, we show the effect of Pd/Pt/Re/Rh alloying on Fe site in CeFe$_{2}$. A monotonic change was observed in phase stability with increasing at.\%Fe. The Fe shows large solubility range from min(67 at.\%) for Ce(Fe$_{z}$Re$_{1-z}$)$_{2}$ in Fig.~\ref{Ce_stability}b$_{2}$, and max (80 at.\%) for Ce(Fe$_{z}$Pd$_{1-z}$)$_{2}$ in Fig.~\ref{Ce_stability}b. However, (near)linear or positive slope in formation enthalpy indicates that Fe does not mix and completely phase separates with other alloying elements. Given the stability of CeFe$_{2}$ in cubic phase, we can understood phase stability in Fig.~\ref{Ce_stability}b-b$_{3}$ in two ways - (i) Ce-Fe-Y may form different stable layered structure, or (ii) Ce-Fe and Ce-Y will phase separate and coexist together. On the other hand, weakly positive E$_{form}$ near Fe-rich region indicates towards possible thermal stabilization of pure binary CeFe$_{2}$ compound. {Notably, the ML predicted E$_{form}$ for CeFe$_{2}$ ($\sim$0.04 eV-atom$^{-1}$) shows reasonably good agreement DFT calculations  (0.06 eV-atom$^{-1}$).}

%The DFT calculated E$_{form}$ for CeFe$_{2}$ (0.06 eV-atom$^{-1}$) shows reasonably good agreement with ML predictions ($\sim$0.04 eV-atom$^{-1}$).

The phase stability of Ce-Mn based rare-earth compound alloyed with Pd/Pt/Re/Rh at Mn-site is shown in Fig.~\ref{Ce_stability}c-c$_{3}$. The Ce-Mn system is more interesting in the sense that CeMn$_{2}$ does not exist. The large positive change in phase stability of Ce(Mn$_{z}$TM2$_{1-z}$)$_{2}$ (TM2=Pd/Pt/Re/Rh) on increasing Mn concentration in Fig.~\ref{Ce_stability}c-c$_{3}$ shows a poor solubility of Mn, which is clear signature of phase change beyond 20-40 Mn at.\%. {The ML predicted phase stability for CeMn$_{2}$ ($\sim$0.19 eV-atom$^{-1}$) is in good agreement with DFT (+0.12 eV-atom$^{-1}$)}, which further affirms the thermodynamic instability of CeMn$_{2}$. Going back to our discussion of energy instability in CeRe$_{2}$, {this result confirms} metastability or dynamically instability of Ce-Re-Mn compounds in MgCu$_{2}$ phase. 

The alloying effect of Ni with 4$d$/5$d$ (TM2=Pd/Pt/Re/Rh) transition metals on the phase stability of Ce(Ni$_{z}$TM2$_{1-z}$)$_{2}$ compounds was also analyzed in Fig.~\ref{Ce_stability}d-d$_{3}$ due to special interest towards RE-Ni compounds for their magnetocaloric properties and hydrogen absorption capacity \cite{Oesterreicher1976,Buschow1972}.
Experimentally, CeNi$_{2}$ is well known to crystallize in cubic Laves phase with lattice parameter of 7.194 \AA~\cite{Wallace1970}. Our model in Eq.~\ref{Eq_eform} also predicts strong stability of CeNi$_{2}$ with E$_{form}$ of  -0.18 eV-atom$^{-1}$ as shown in Fig.~\ref{Ce_stability}d-d$_{3}$. The predicted stability was found in good agreement with the DFT (-0.24 eV-atom$^{-1}$) and experimentally observed cubic MgCu$_{2}$ crystal structure \cite{Wallace1970}. 

In Fig.~\ref{Ce_stability}d, the chemical alloying of Ni with Pd in CeNi$_{2}$ shows a weak effect on phase stability up to 40 at.\%Ni, while a sharp increase in stability was found in Ni-poor region. Model predictions show a very high stability for CePd$_{2}$ (-0.29 eV-atom$^{-1}$), which is in good agreement with the DFT (-0.24 eV-atom$^{-1}$). Although CePd$_{2}$ was never reported  experimentally, our calculations suggest its possible existence in MgCu$_{2}$ crystal phase. Our model also predict the stability of CePt$_{2}$ as shown in Fig.~\ref{Ce_stability}d$_{1}$, which is in agreement with experimental observation of cubic MgCu$_{2}$ crystal structure \cite{Villars2012}. However, the near linear trend in E$_{form}$ suggests that Ni and Pt either mix very weakly or phase-separate. 

In Fig.~\ref{Ce_stability}d$_{2}$, the phase stability analysis of CeRe$_{2}$ shows very interesting behavior, where alloying with Re weakens the phase stability of CeNi$_{2}$ (from 100-70 at.\%Ni or 0-70 at.\% Re), while Re-rich (Ni-poor) region shows increased phase stability. A weakly unstable region for 20-52 at.\%Ni is also interesting because of transition from energetically `stable' to `unstable' region. The discrete phase stability in Ce-Ni-Re is suggestive of possible phase transformation, which needs more careful exploration both from theory and experiments as binary Ce-Re compounds are yet to be realized.  

The Ce-Rh based compounds, on the other hand, are more interesting because of their critical physical behaviors such as Kondo-effect and Fermi-surface effects~\cite{Higuchi1994}. Therefore, we also assessed the effect of Rh alloying on thermodynamic stability of CeNi$_{2}$ in Fig.~\ref{Ce_stability}d$_{3}$. The model predicts improved stability of CeRh$_{2}$ (-0.60 eV-atom$^{-1}$) with respect to CeNi$_{2}$, which is in good agreement with DFT (E$_{form}$(CeRh$_{2}$)=-0.72 eV-atom$^{-1}$). The experimental realization of CeRh$_{2}$ \cite{Sugawara1994} further affirms the robustness of our prediction. However, the linear trends for Ce-Ni-Rh in E$_{form}$ again suggests that Ni and Rh may not want to mix or mix very weakly. However, strong energy stability suggests that both the phases, i.e., CeNi$_{2}$ and CeRh$_{2}$, may co-exist together.

\subsection*{Effect of chemical mixing of 3$d$ transition metals on phase stability}

\begin{figure}[htp]
    \centering
    \includegraphics[width=0.85\textwidth]{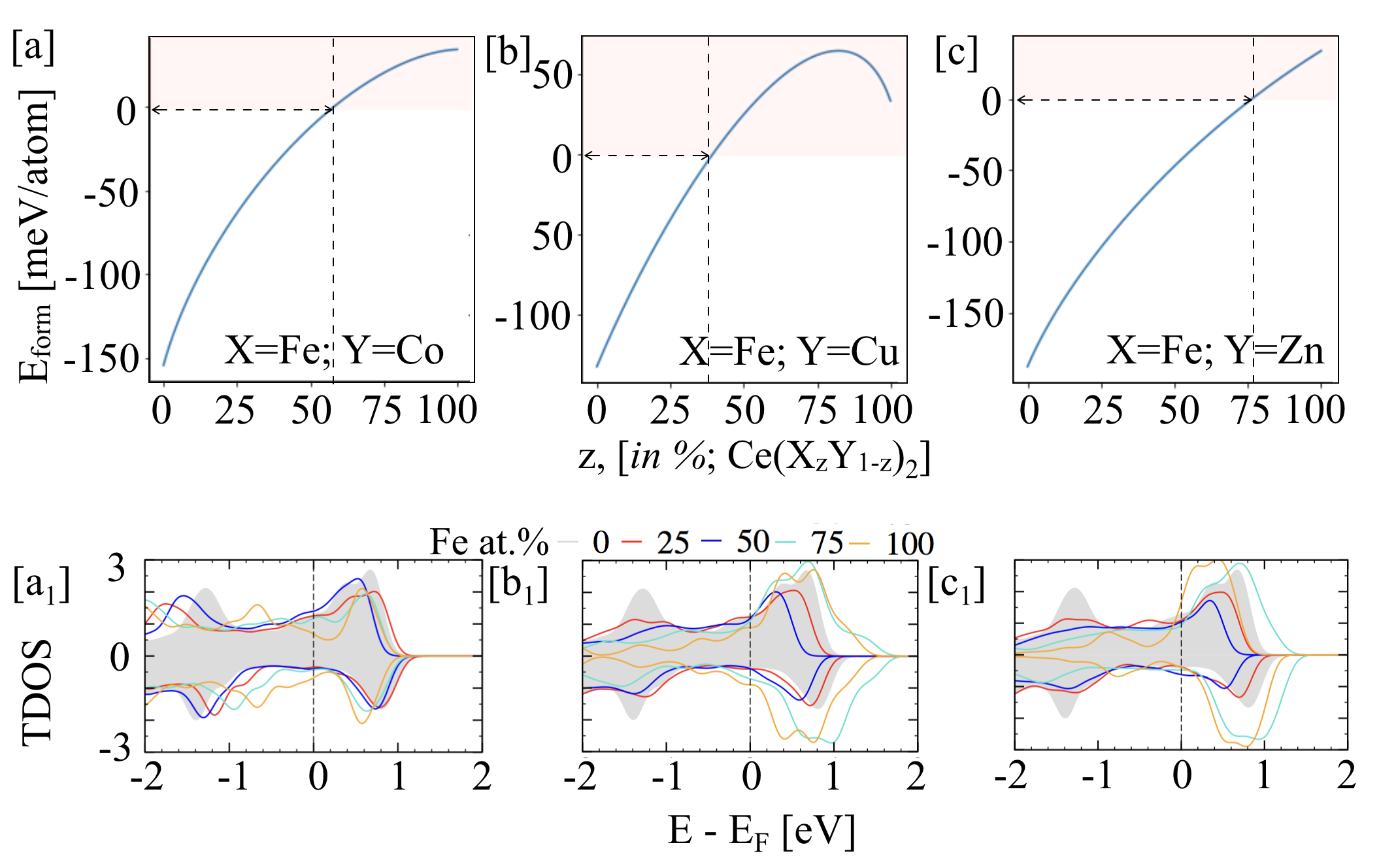}
    \caption{(Color Online) (a-c) Chemical alloying of chemically dissimilar atoms on phase stability of cerium based pseudo-binary compounds Ce(Fe$_{z}$TM2$_{1-z}$)$_{2}$, where TM2=Co/Cu/Zn. (a$_1$-c$_1$) Total density of states at stoicheometric compositions, i.e., z (Fe)=0, 25, 50, 75, and 100 at.\%. The solubility range of Co/Cu/Zn at Fe is marked by arrow.}
    \label{Ce_stability2}
\end{figure}

\begin{figure}[htp]
    \centering
    \includegraphics[width=0.85\textwidth]{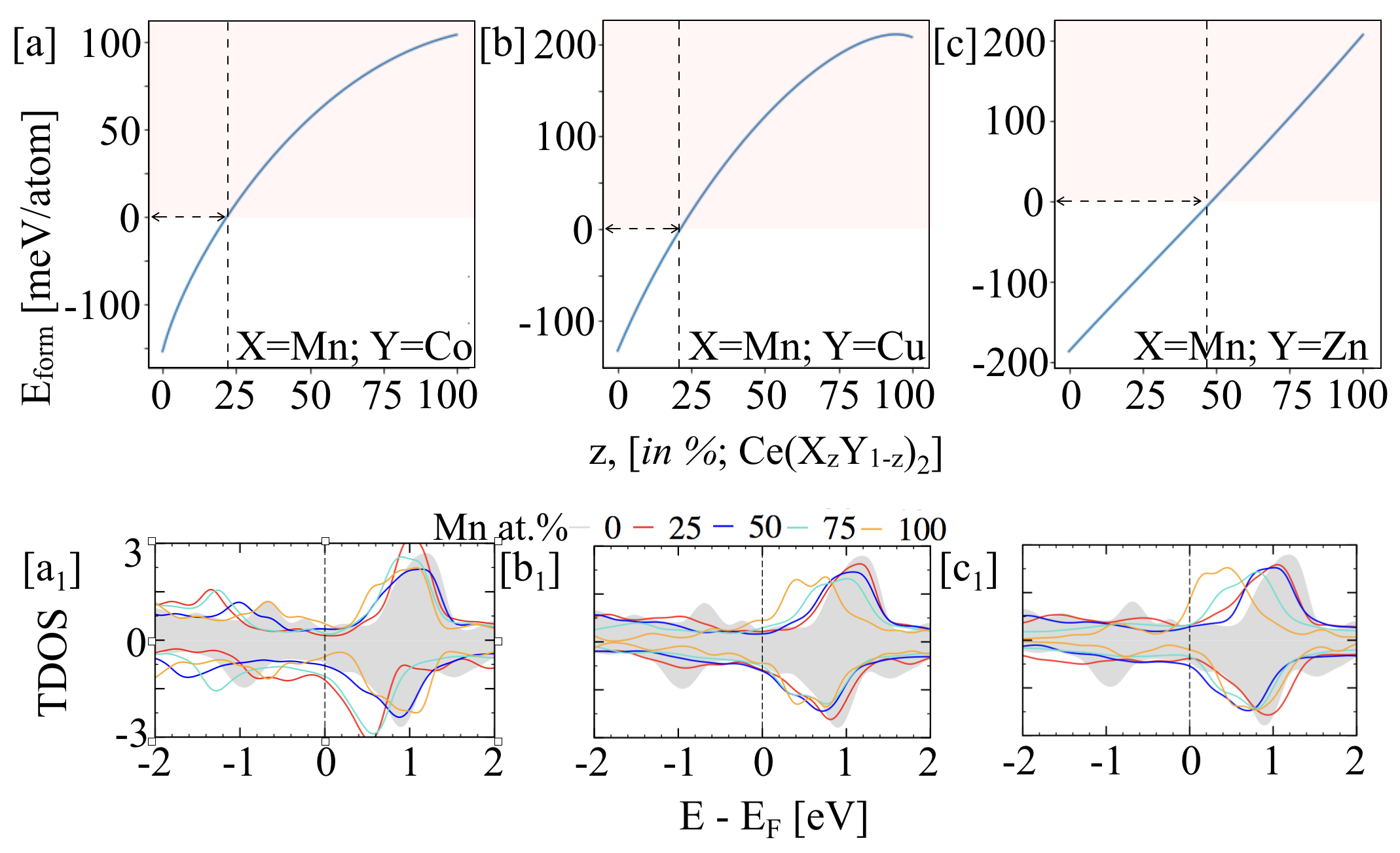}
    \caption{(Color Online) (a, c) Effect of alloying on the phase stability of cerium based pseudo-binary compound Ce(Mn$_{z}$TM2$_{1-z}$)$_{2}$ with chemically distinct atoms (with similar atomic-radii), where TM2=Co/Cu/Zn. (a$_1$-c$_1$) Total DOS at stoichiometric compositions, i.e., z (Mn)=0, 25, 50, 75, and 100 at.\%. The solubility range of Co/Cu/Zn at Mn is marked by arrow.}
    \label{Ce_stability3}
\end{figure}

Low-cost rare earths such as Ce are becoming a popular topic for future applied research due to the criticality of currently available high-performance permanent magnets \cite{Zhou2013}. Therefore, it would be useful to explore the thermodynamic stability of Ce based compounds, which contain earth-abundant elements and may form the base for future hard magnetic materials. While the MgCu$_{2}$ type structure is isotropic and unlikely to support hard ferromagnetism, it is useful to see how alloying by common 3d metals can create lattice distortions producing uniaxial symmetry. In Fig.~\ref{Ce_stability2}a-c, we mixed chemically dissimilar 3$d$ transition elements with similar atomic-sizes together to understand the effects of electron doping (change in valence-electron count) on  thermodynamic stability of Ce(TM1$_{z}$TM2$_{1-z}$)$_{2}$, where TM1=Fe/Mn; and TM2=Co, Cu, and Zn. X=Fe was chosen for its robust magnetic behavior, while Mn was chosen for its intriguing and complex magnetic character. In Fig.~\ref{Ce_stability2}a-c$_{1}$ and Fig.~\ref{Ce_stability3}a-c$_{1}$, we analyze the  effect of chemical mixing of TM1=Fe/Mn with TM2=Co/Cu/Zn on phase-stability and  electronic-structure of Ce(TM1$_{z}$TM2$_{1-z}$)$_{2}$. {The CeFe$_{2}$ is stable in MgCu$_{2}$-type phase at room temperature, however, in its ground state it shows a rhombohedral distortion \cite{Haldar2010}. Whereas, no reports were found that show the existence of cubic structure in Mn based Laves phases. Thus, the explored scenarios are largely hypothetical, allowing us to look into thermodynamic stability of experimentally unavailable rare earth phases using machine learning models.} 

The ML-predicted phase stability in Fig~\ref{Ce_stability2}a-c indicates a Fe solubility limit  of 62/37/77 at.\% in Ce(Fe$_{z}$TM2$_{1-z}$)$_{2}$, when chemically mixed with Co/Cu/Zn alloying elements, respectively. This indicates that Fe in Ce(Fe$_{z}$TM2$_{1-z}$)$_{2}$ is weakly stable (or unstable) in cubic phase beyond  62/37/77at.\%Fe. The mixing of Co/Cu/Zn at Fe is significant as each of the alloying elements stabilizes the pure CeFe$_{2}$ phase as shown in Fig~\ref{Ce_stability2}a-c. An important characteristic was observed in Fig.~\ref{Ce_stability2}c, where the formation enthalpy curve is nearly straight line. This can be attributed to chemical inactivity of the nearly half-filled Fe-3$d$ (6e$^{-s}$s) and completely filled Zn-3$d$ (10e$^{-}$s) shells. The predicted trends in stability of the end compounds, i.e., CeFe$_{2}$ (DFT= 4 meV-atom$^{-1}$; ML= 20 meV-atom$^{-1}$) and CeTM2$_{2}$ (TM2=Co/Zn; DFT=($\approx$-0.200/-0.250 meV-atom$^{-1}$); ML=-0.152/-0.190 meV-atom$^{-1}$), compares well with the DFT numbers. For CeCu$_{2}$, the ML model  predicts formation enthalpy of -0.135 eV-atom$^{-1}$ in cubic phase, which is thermodynamically less stable than the orthorhombic (-0.175 eV-atom$^{-1}$) phase. The ML predictions are is agreement with both DFT and experiments. 

To shed more light on trends in energy stability, in Fig.~\ref{Ce_stability2}a$_1$-c$_1$, a zoomed-region in total DOS near the Fermi-level  of Ce(Fe$_{z}$TM2$_{1-z}$)$_{2}$ (TM2=Co/Cu/Zn) is shown. The stability at low to intermediate (0-50 at.\%) Fe concentration in Fig.~\ref{Ce_stability2}a is attributed to well-structured DOS peaks just below the Fermi-level, which subsequently disappears on further alloying. Furthermore, Co and Fe bands show no overlap, employing weak or no hybridization. Similarly, the Fe solubility range in Fig.~\ref{Ce_stability2}b drops drastically to near 30 Fe at.\% with Cu alloying whereas it increases to 77 at.\% for Zn alloying in Fig.~\ref{Ce_stability2}c. Despite increasing stability due to Fe/Zn alloying, the Fe/Zn DOS shows no hybridization, i.e., Zn-3$d$ and Fe-3$d$ are energetically far-apart. 

The underlying magnetic structure also plays an important role on thermodynamic stability. As pointed out by Eriksson \etal \cite{Eriksson1988} in CeFe$_{2}$, it is the itinerant nature of electrons in the Ce$-4f$ and Fe$-3d$ states that are coupled anti-ferromagnetically (AFM). The AFM structure stabilizes CeFe$_{2}$ in the cubic phase. The hybridization between Ce-4$f$ states with Fe-3$d$ itinerant electronic systems can be controlled by compositional tuning. To contrast between itinerant and localized electrons, we choose CeCu$_{2}$ as an example where the alloy stabilizes in orthorhombic phase, where the Ce moments are anti-ferromagnetically arranged~\cite{Nunezt1992}. This is different from CeFe$_{2}$, where Ce and Fe moments are anti-parallel, which suggests that itinerant electrons hybridize differently than localized 4$f$ electrons in CeCu$_{2}$. The stability of given phases can also be correlated to the valence-electron count (VEC) of Fe-3$d$ (VEC=8) and Cu-3$d$ (VEC=11). This electronic dissimilarity between Fe and Cu explains the stability of CeFe$_{2}$ in cubic phase and CeCu$_{2}$ in orthorhombic phase. The relation of phase stability with electron-count has also been explained by Gschneidner \etal~\cite{Gschneidner2006} using the idea of an electron-per-atom ratio.

Thermodynamic stability of Ce(Mn$_{z}$TM2$_{1-z}$)$_{2}$, where Mn is alloyed with TM2=Co/Cu/Zn, is shown in Fig~\ref{Ce_stability3}a-c. The solubility range for of Mn is marked by a dashed arrow. We observed that the predicted formation enthalpy remains positive for most Mn-rich composition range. The solubility limit of Mn in Ce(Mn$_{z}$TM2$_{1-z}$)$_{2}$ is significantly reduced (23/23/50 Mn at.\%) compared to the Fe-based compounds (62/37/77 at.\%), when alloyed with Co/Cu/Zn (TM2). Intuitively, the phase stability predictions in Fig.~\ref{Ce_stability3}c for Mn mixed with Zn show linear (straight-line) change, which suggests that Zn does not want to mix with Mn. Notably, the reason for this trend can also be understood from individual electronic configurations of Mn, which has half-filled 3$d$ bands (i.e., 5e$^{-}$'s), while Zn remains chemically-inactive due to its completely filled $d$-bands (10e$^{-}$'s in 3$d$), i.e., alloying of Mn with Zn have weak or no chemical activity. Also, the positive E$_{form}$ for CeMn$_{2}$ agrees with experiments, as this alloy is unstable in the MgCu$_{2}$ crystal structure. On the other hand, the CeFe$_{2}$ is a well known phase synthesized experimentally in MgCu$_{2}$ structure, however, the weak stability of CeFe$_{2}$ both in ML and DFT predictions can be attributed to relatively large density of states peak (in orange at 100 at.\%Fe; and cyan 75 at.\%Fe in Fig.~\ref{Ce_stability2}) near the Fermi-level. It is worth mentioning that the low-T and room-temperature structures of CeFe$_{2}$ are different, and the structural transition is possibly driven by magnetism. On the other hand, in almost all cases with Fe-poor regions, the increased stability is attributed to a pseudo-gap at E$_{Fermi}$ in the total DOS. Similar behavior was also found in more complex alloy systems~\cite{Singh2018}.

\subsection*{Experimental validation of the {machine-learning predicted} phase stability of Ce-Fe-Cu compounds}

\begin{figure}[htp]
\centering
\includegraphics[width=1\textwidth]{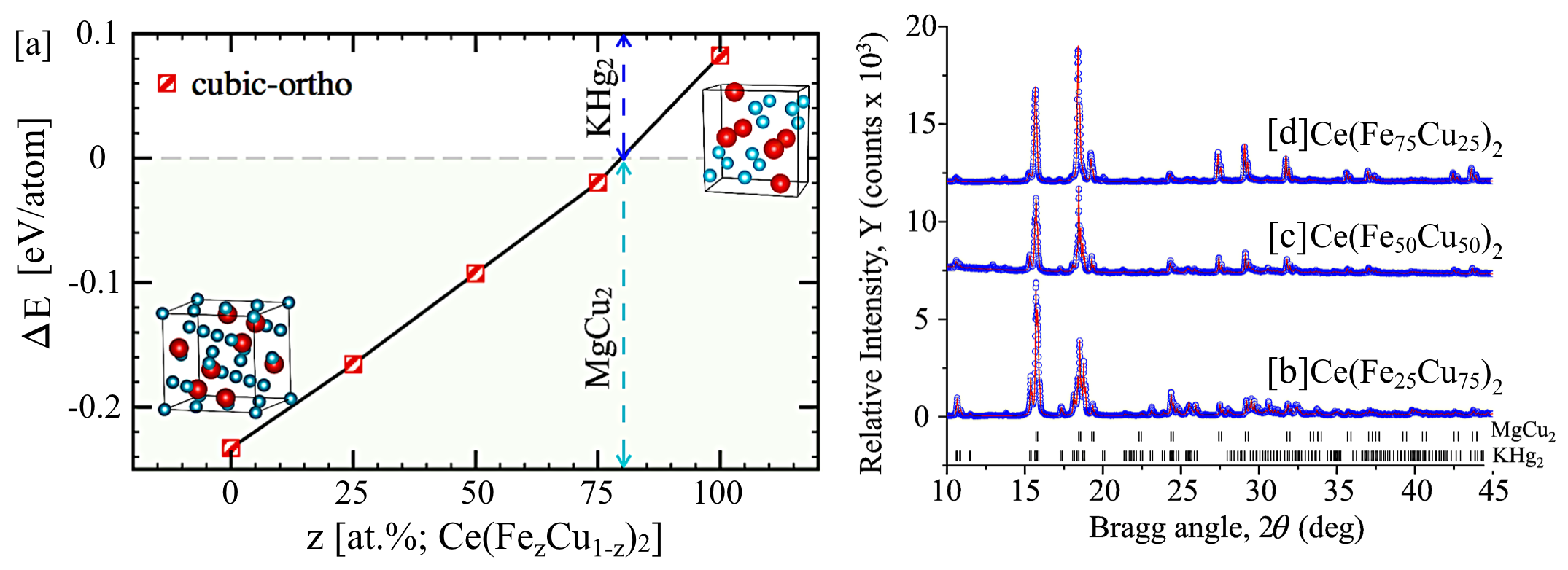}
\caption{(Counter-clockwise) (a) The energy difference ($\Delta{E}$) between cubic and orthorhombic phase was plotted for Ce(Fe$_{z}$Cu$_{1-z}$)$_{2}$. (b-d) The X-ray powder diffraction pattern at z = 25, 50, and 75 at.\%Fe in Ce-Fe-Cu compound shows the presence of mixed cubic and orthorhombic phases.}
\label{CeFuCu}
\end{figure}

In Fig.~\ref{CeFuCu}a, we plot DFT calculated phase energy difference ($\Delta{E}$; eV-atom$^{-1}$) for Ce(Fe$_{z}$Cu$_{1-z}$)$_{2}$ comparing cubic (MgCu$_{2}$) and orthorhombic (KHg$_{2}$) phases with respect to the end points (CeFe$_{2}$ and CeCu$_{2}$). The CeFe$_{2}$ is known to stabilize in cubic phase, whereas CeCu$_{2}$ in orthorhombic phase. The ML predicted phase stability of cubic and orthorhombic phases of CeCu$_{2}$  is -0.137 eV-atom$^{-1}$ and -0.175 eV-atom$^{-1}$, respectively, which is in agreement with the experiments. Similarly, DFT-calculated $\Delta{E_{form}}$ in Fig.~\ref{CeFuCu}a shows that cubic phase is stable at 0at.\%Cu (100at.\%Fe) while orthorhombic phase is preferable at 100at.\%Cu, in agreement with experiments \cite{Roy2004,Seyfarth2021}. Phase stability in Fig.~\ref{CeFuCu}a shows a crossover from cubic to orthorhombic phase at 80at.\%Cu. Importantly, the thermodynamic stability could be controlled using compositional variation at transition metal site in CeTM$_{2}$, TM=Fe or Cu.  A detailed analysis of phase stability was performed through band-structure, charge-density and magnetization density in supplemental Fig.~S4\&~S5.

To validate our predictions, we synthesize the Ce-Fe-Cu compounds at x (Fe)=25, 50 and 75 at.\%. Our XRPD data in Fig.~\ref{CeFuCu}b-d further confirms our predictions that CeFe$_{2}$ and CeCu$_{2}$ do not form a solid solution rather they energetically prefer to exist as a mixed cubic (CeFe$_{2}$) and orthorhombic (CeCu$_{2}$) intermetallic phase. The ratios of CeFe$_{2}$ and CeCu$_{2}$ in the prepared alloys, as determined by Rietveld refinement, agree with the nominal Fe/Cu ratios, indicating that in studied compositions the alloying simply form two-phase mixture avoiding chemical interaction. The Cu atomic position in the CeCu$_{2}$ compound (KHg$_{2}$ structure) does not take any Fe and the lattice parameters of this phase do no change significantly (i.e. the change is $<$0.01~\AA). The behavior of CeFe$_{2}$ is more interesting since both EDX and XRPD refinement indicate that a small amount of Cu (EDS suggests $\sim$3\%) is likely to mix with Fe in the transition metal position. This is in agreement with calculations that predict lower energy with Cu addition. However, the lattice parameter a of prepared alloys, obtained in the range of 7.308$\pm$0.003 (z(Fe)=25) to 7.314$\pm$0.003~\AA~(z(Fe)=75), is higher than that reported for pure CeFe$_{2}$  (typically reported as $\sim$7.30~\AA). This is in odds with the fact that atomic radii of Cu is slightly smaller or nearly equal to that of Fe. One may speculate that Cu presence slightly decreases Ce valence correspondingly increasing its atomic radii. A number of interesting behaviors has already been reported in lightly doped CeFe$_{2}$  compounds, so, perhaps, a closer look is warranted at the Ce(Fe$_{z}$Cu$_{1-z}$)$_{2}$ system, where z$<$0.03, as well. In regard to this work, the experimental data clearly support the theoretical results of restricted mixing of Fe and Cu atoms in the Ce(Fe$_{z}$Cu$_{1-z}$)$_{2}$ system.

\subsection*{Electronic-structure origin of the phase stability of Ce-Fe-Cu compounds and correlation with machine-learning descriptor}

\begin{figure}[htp]
\centering
\includegraphics[width=1\textwidth]{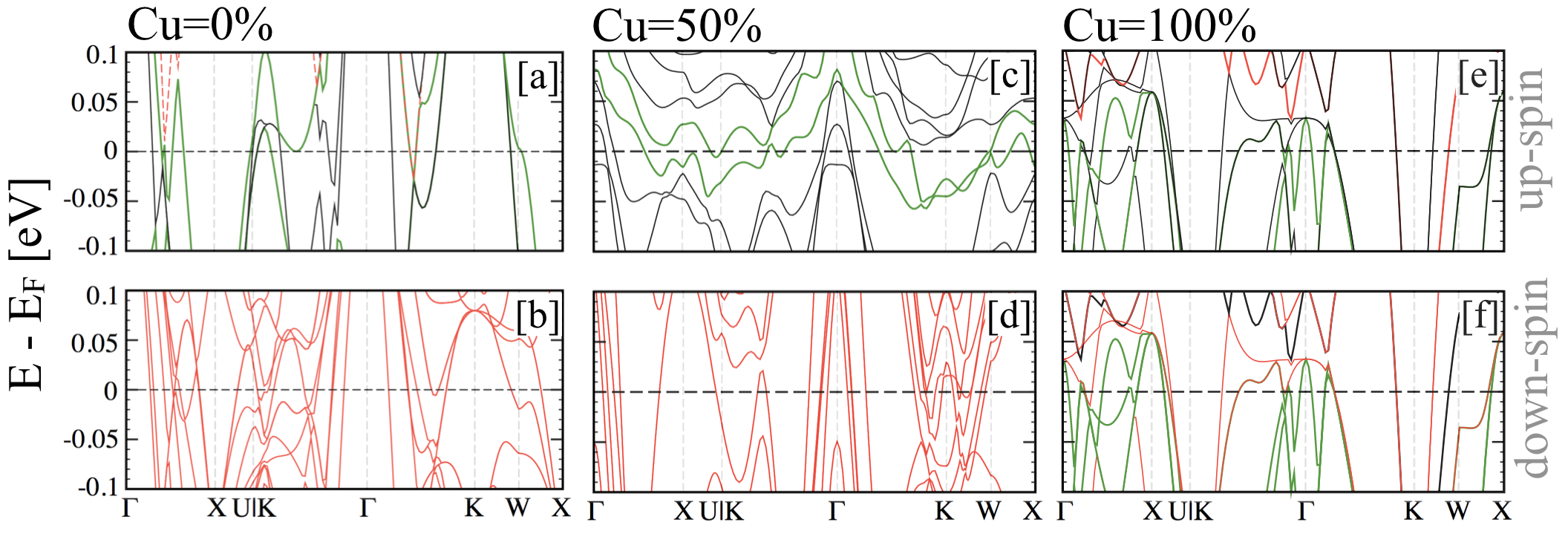}
\caption{(Color Online) The spin-polarized band-structure of (a,b) CeFe$_{2}$, (c,d) Ce(Fe$_{0.5}$Cu$_{0.5}$)$_{2}$, and (e,f) CeCu$_{2}$ in cubic phase of Ce-Fe-Cu based rare-earths. The top- and bottom-panels represent the up-spin and down-spin, respectively.}
\label{CeFuCu_bs}
\end{figure}

\begin{figure}[htp]
\centering
\includegraphics[width=0.7\textwidth]{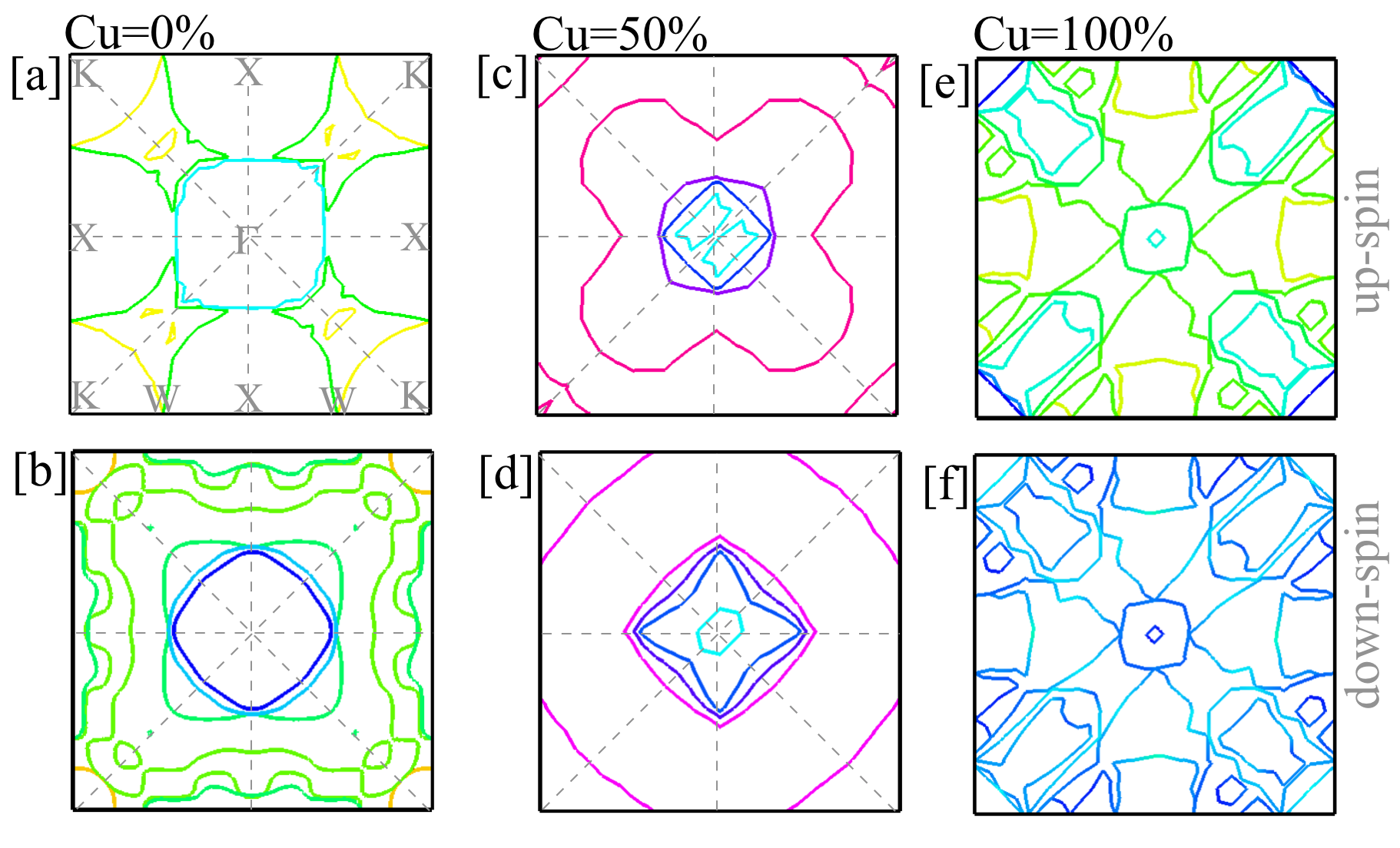}
\caption{(Color Online) The (001) cross-section of the Fermi-surface (at E$_{Fermi}$=0) for (a,b) CeFe$_{2}$, (c,d) Ce(Fe$_{0.5}$Cu$_{0.5}$)$_{2}$), and (e,f) CeCu$_{2}$. The top- and bottom-panels represent the up-spin and down-spin, respectively. Band decomposed Fermi surfaces are shown in Fig.~S6, S7, and S8.}
\label{CeFuCu_fs}
\end{figure}

{The appearance of atomic-features in the E$_{form}$ descriptor in Eq.~\ref{Eq_eform} in the context of phase stability has already been discussed in the feature analysis section. However, it will also be critical to understand the electronic-structure origin of phase stability and its correlation with the E$_{form}$ descriptor. Therefore, we analyzed the DFT calculated band-structure of Ce-Fe-Cu compounds in Fig.~\ref{CeFuCu_bs}~\&~\ref{CeFuCu_fs}.} In Fig.~\ref{CeFuCu_bs}, we show the spin-polarized band-structure of Ce(Fe$_{z}$Cu$_{1-z}$)$_{2}$ at x=0, 50, 100at.\%Cu. The CeFe$_{2}$ and Ce(Fe$_{0.50}$Cu$_{0.50}$)$_{2}$ band-structures are shown along high-symmetry directions of cubic phase, while the bands-structure of CeCu$_{2}$ is shown along high-symmetry directions of orthorhombic phase in the energy-range (E$-$E$_{Fermi}$) -0.1 to 0.1 eV. The highlighted bands in the band-structure are Ce$-4f$ states that are localized near the Fermi-level both in the cubic (Fig.~\ref{CeFuCu_bs}a-d) and orthorhombic (Fig.~\ref{CeFuCu_bs}e,f) phases. The DFT calculated magnetic moment-contribution of each species at 0 at.\%Cu in the cubic phase is  (Ce=-0.68, Fe=1.98)~$\mu_{B}$, whereas magnetic moment of individual species changes to (Ce(1) =0.55(near Cu), Ce(2) = 0.61 (near Fe); Fe (1) =-1.61, Fe(2)=-1.42;  Cu=0.03)~$\mu_{B}$ at 50 at.\%Cu. The change in magnetic behavior shows the effect of change in neighboring environment of Ce and Fe sites on alloying with Cu. Notably, we found that both Ce and Fe moments drop with increasing at.\%Cu, and the moments drop to zero  in the cubic phase where Fe=Cu, i.e., Ce=0; Cu=0 $\mu_{B}$. The reduced magnetic character is directly correlated with the decreased thermodynamic stability of the  Ce-Fe-Cu compound. Moreover, the magnetic moment at the Ce/Fe sites in the cubic phase were flipped by adding Cu at the Fe-site in CeFe$_{2}$. The increasing Cu concentration increases the cell volume in the cubic phase, which  was in agreement with the experimental data shown above and  found to be responsible for decrease in both magnetic character and phase stability of Ce-Fe-Cu alloy. As can be seen from the band-structure in Fig.~\ref{CeFuCu_bs}e,f, the magnetic character changes from strong-ferromagnetic (CeFe$_{2}$) to non-magnetic (CeCu$_{2}$) in the cubic phase.

The energetically stable orthorhombic phase at 100at.\%Cu in Fig.~\ref{CeFuCu} has a finite Ce$-4f$ moment of 0.60 $\mu_{B}$, however, Ce$-5d$ has lost its magnetic behavior in absence of Fe$-3d$ states (see comparative plot of cubic and orthorhombic phases at 100at.\% in supplement Fig.~S2). The decrease in phase stability of the cubic phase in Fig.~\ref{CeFuCu} strongly correlates with increasing at.\%Cu, as shown in the total DOS plot in Fig.~\ref{Ce_stability2}b$_{1}$. The electronic density at E$_{Fermi}$ was found to increase with increasing at.\%Cu, which is evident from the presence of a 'spaghetti' configuration of bands in the spin-polarized band-structure plot at 50 at.\%Cu in Fig.~\ref{CeFuCu_bs}c,d. The behavior of Ce$-4f$ states changes from wavy (represent more hybridization) to much flatter representing reduced hybridization between Ce-Cu due to filled Cu$-3d$ states at 50~at.\%Cu.

{Microscopically, the delocalization degree of valence electrons can significantly impact the magnetic behavior. Therefore, the loss of magnetic character in Cu doped CeFe$_2$ could also be connected to the difference in electronegativity and atomic-radii of Fe ($\chi$=1.80, r=1.40 \AA) with respect to Cu ($\chi$=1.85, r=1.35 \AA), especially since higher electronegativity correlates with lower radii due to stronger attraction between the nucleus and valence electrons. Also, Li et al. \cite{Li2012} has shown that electronegativity and atomic-size difference play an important role in influencing the magnetic behavior by modifying attractive interaction between valence electrons of atoms. Looking back to Eq.~\ref{Eq_eform} and correlating that to electronic-structure analysis, the appearance of electronegativity and atomic-radii in E$_{form}$ descriptor becomes more understandable. While training or cross-validation of the SISSO model was provided with no biases, still the final E$_{form}$ descriptor picks important physical quantities such as atomic-radii, electronegativity which directly connects to phase stability. This discussion is also consistent with our electronic-structure analysis, which further signifies the importance of using the physics-based descriptors for materials design.}

We show the (001) projected Fermi-Surface in the cubic phase of Ce(Fe$_{z}$Cu$_{1-z}$)$_{2}$ at 0, 50, and 100 at.\%Cu in Fig.~\ref{CeFuCu_fs}a-f. Our aim was to understand stability in terms of change in complexity of the Fermi-surface. Is there any relation? In Fig.~\ref{CeFuCu_fs}a,b, the (001) Fermi-surface cross-section of CeFe$_{2}$ shows hole-pockets at $\Gamma$ and K points, and electron-pockets at the X-point in both up-and down-spin channels. Down-spin-channel has multiplet of hole-pockets mainly arising from Fe-$3d$ states while hole pockets in the up-spin channel are mainly from Ce$-4f$. A similar characteristic was found in band plot in Fig.~\ref{CeFuCu_bs}a,b. We found that Cu alloying drastically changed the band-structure in Fig.~\ref{CeFuCu_bs}c,d, therefore, it would be interesting to see how Cu affects the Fermi-surface. The (001) Fermi-surface cross-section is shown in Fig.~\ref{CeFuCu_fs}c,d, where two hole-pockets were found in the up-spin channel while four hole-pockets in the down-spin channel at $\Gamma$-point.  The hole-pocket got much bigger at X-point in down-spin channel in Fig.~\ref{CeFuCu_fs}d, while hole pocket in up-spin channel at X-point disappears. This characteristic was seen in band-structure due to change in magnetization direction CeFe$_{2}$ with Cu alloying. In Fig.~\ref{CeFuCu_fs}e,f, CeCu$_{2}$ becomes non-magnetic where X and $\Gamma$ show much smaller hole-pockets (one each) in both up-and down-spin channels. On the other hand, a Fermi-surface cross-section along (001) shows a slightly enlarged electron-pocket at K-point. This is in agreement with the fact that Cu alloying adds excess electrons. The reduction of hole-pockets on Cu alloying indicates filling of all bonding states, which drastically reduces the hybridization of Ce with transition metals and reduces stability of Ce-Fe-Cu pseudo-binary alloys in cubic phase.

{\it {Generalizability of the SISSO based analytical models:~}} {Regarding generalizability of the model and its ability to go beyond known chemistries, we want to emphasize that none of the predicted disordered cases in Fig.~\ref{Ce_stability}-\ref{Ce_stability3} were included in the training or test data. Our calculations show that analytical models are useful for predictive design of disordered alloys.  Furthermore, {\it ab-initio} methods struggle with this because the design of non-stoichiometric disordered supercells is non-intuitive and computationally expensive. In the case where one of the end points is known to not exist, a solubility limit can still be obtained (e.g., solubility limit of Fe in Cu if CeFe$_{2}$ did not form). Our model predictions for Ce-Fe-Cu shows that Fe-Cu do not want to mix, which is independently verified by DFT calculations and experiments in Fig.~\ref{CeFuCu}. These findings are promising as they tentatively suggest the model is capable of making safe extrapolations.}

{We also want to point out that SISSO may not be the only or even best modeling framework that could be used to carry out the proposed task. We note that a priori, it is not always immediately clear when a model is sufficiently simple and yet expressive enough to carry out robust predictions. We highlight, for example, the work by Bartel \etal~on the use of SISSO to predict finite temperature corrections to the Gibbs energies of ordered compounds: at the end, the SISSO model is relatively simple (a posteriori), but by carrying out the SISSO methodology the authors were able to explore (order) billion-dimensional feature spaces, arriving at the conclusion that the best model indeed was highly parsimonious, and yet, highly predictive \cite{Bartel2018}.}

%%%%%%%%%%%%%%%%%%%%%%%%%%%%%%%%%%%%%%
\section*{Conclusion}
%%%%%%%%%%%%%%%%%%%%%%%%%%%%%%%%%%%%%%

Physical descriptor-based analytical machine-learning models are very useful for accelerated identification of new inorganic compounds with varying chemistry. However, the absence of sufficient density-functional theory or experiment generated entries into the rare-earth databases makes the execution of high-quality machine learning extremely difficult, given the fact that generating accurate DFT or experimental database for rare-earths is very tricky as well as time and resource consuming. In this work, we generated nearly 600+ data points for REX$_{2}$ type rare-earth compounds. The SISSO models were trained over a sparse `in-house' database that provides three dimensional (3D) analytical descriptor. We demonstrated that computationally inexpensive descriptors can be incorporated to predict the equilibrium phases of rare-earth based compounds purely on a thermodynamics basis, i.e., using formation enthalpies. We present a detailed phase stability analysis of Ce based rare-earth compounds with cubic Laves phase and provide a quantitative guidance for compositional considerations in realizing new metastable materials with unconventional chemistries. We validated our ML and DFT predictions of phase stability through very careful synthesis of Ce(Fe$_{z}$Cu$_{1-z}$)$_{2}$ compounds with 25, 50 and 75 at.\%Fe(z). Our experiments show that Fe and Cu do not form solid solution rather they stabilize as mixed intermetallic phases, i.e., CeFe$_{2}$ and CeCu$_{2}$, which is in good agreement with model predictions. The low computational cost and reasonable accuracy of interpretable machine-learning models presented in this work will enable the high-throughput analysis of all possible crystal phases and remove external biases in phase selection. Despite the successful use of limited database and predicting the stability of rare-earth compounds and mixed compositions, there are a number of other challenges that need to be addressed, for example, (i) temperature, and (ii) pressure dependence of ground state structure and its impact on thermodynamic stability. While extracting the complete picture using standard DFT calculations would be prohibitively challenging, we believe that ML methods described here will enable such evaluations and bring the science closer to the desired "materials from PC" land. Our approach will be useful in discovering new and complex rare-earth compounds with new functionalities.

{Furthermore, more insights are probably needed into the extrapolation of the SISSO descriptors and the chosen featurization for the complete generalization of descriptors beyond training and test data. However, this requires a separate, yet a detailed investigation to ascertain the ability of this model to extrapolate, with reasonable accuracy, beyond rare-earth and transition metal based alloying chemistries. Notably, simple predictive approaches, such as a second-order polynomial, require unknown coefficients with no physical meaning would still require the utilization of either regression or other machine learning approaches. This suggests that there is no good way to get around the machine learning aspect. Notably, SISSO provides an interpretable analytical model that makes its simpler to use for predictive design, while other regression techniques are mostly a black box. At the same time, there is clearly a need for more insightful investigation into the extrapolation of the SISSO descriptors for arbitrary alloying chemistries as machine-learning are primarily data sensitive.} 

\section*{Acknowledgements}
P.S. and Y.M. acknowledge the insightful discussions with Dr. Vitalij K. Pecharsky and Dr. Duane D. Johnson of Ames Laboratory at Iowa State University. This work was supported by Laboratory Directed Research and Development Program (LDRD) program at Ames Laboratory. Work at Ames Laboratory was supported by the U.S. Department of Energy (DOE), Office of Science, Basic Energy Sciences, Materials Science \& Engineering Division. Ames Laboratory operated by Iowa State University for the U.S. DOE under contract DE-AC02-07CH11358. GV and RA acknowledge the support of QNRF under Project No. NPRP11S-1203-170056. DFT calculations were carried out at the Texas A\&M High-Performance Research Computing (HPRC) Facility.

\section*{Data availability}
The authors declare that the data supporting the findings of this study are available within the paper and supplement. Also, the data that support the plots within this paper and other findings of this study are available from the corresponding author upon reasonable request.

\section*{References}
\biboptions{sort&compress}

\end{document}